\documentclass[9pt,twoside]{rmaa-rho-class/rmaa-rho}
\makeatletter
\let\linenumbers\relax

\makeatother
\RMxAAtemplatetype{\RMxAA} 

\vol{100}
\pages{1000-1025}
\thisyear{2026}
\doi{\href{https://doi.org/10.22201/ia.01851101p.20XX.XX.XX.XX}{https://doi.org/10.22201/ia.01851101p.20XX.XX.XX.XX}}

\newcommand{\cm}{{~\rm cm}}
\newcommand{\km}{{~\rm km}}
\newcommand{\s}{{~\rm s}}

\newcommand{\g}{{~\rm g}}

\newcommand{\erg}{{~\rm erg}}

\title{Long-lived intermittent accretion disks in the jittering jets explosion mechanism (JJEM) of core-collapse supernovae}

\author[1,$\dagger$]{Noam Soker \orcidlink{0000-0003-0375-8987}}

\affil[1]{Department of Physics, Technion - Israel Institute of Technology, Haifa, 3200003, Israel}

\leadauthor{Noam Soker}
\smalltitle{Long-lived accretion disks in the JJEM}

\corres{Noam Soker}
\email{soker@technion.ac.il}

\received{XXX}
\accepted{\today}

\license{Texto de la licencia aquí}

\setbool{rho-abstract}{true} 
\setbool{rho-resumen}{true} 

\begin{abstract}
Motivated by observations of core-collapse supernova (CCSN) remnants that suggest cases where one to three energetic pairs of jets dominate the CCSN remnant morphology and, hence, the CCSN explosion energy, I examine the formation of long-lived intermittent accretion disks that launch such pairs of energetic jets in the framework of the jittering-jets explosion mechanism (JJEM). In the JJEM, pairs of jets explode all CCSNe. In most CCSNe, stochastic angular momentum fluctuations in the convective zones of the pre-collapse core seed instabilities above the newly born neutron star that lead to the formation of intermittent accretion disks. These disks launch several to about twenty pairs of jets that explode the star. CCSNRs with signatures of 1-3 very energetic pairs of jets require long-lived, intermittent accretion disks. I quantitatively show that viscosity-driven angular momentum transport in the disk can prolong its lifetime even when material with zero angular momentum continues to feed the disk. Other effects that I do not study here can also prolong the disk lifetime somewhat: jets might prevent matter from accreting from the polar direction, and angular momentum fluctuations can, in some cases, add up to a positive angular momentum. The viscosity mechanism I study here, along with these other effects, can prolong the lifetime of 1-3 intermittent accretion disks (or none), which then launch energetic jet pairs. This study adds to the wide variety of morphologies that the JJEM can explain, somewhat supporting the claim that the JJEM is the primary explosion mechanism of CCSNe. 
\end{abstract}


\begin{resumen}
\end{resumen}

\begin{document}

\maketitle
\pagestyle{fancy}\thispagestyle{firststyle}


\section{Introduction} 
\label{sec:intro}

The jittering jets explosion mechanism (JJEM) attributes the explosion of most, and possibly all, core-collapse supernovae (CCSNe) to pairs of jets launched by the newly born neutron star (NS) or black hole (e.g., \citealt{KlimovSoker2026, ShishkinSoker2026SNR0540, Soker2026SN1987Amulecular, Soker2026SNRJ0450, Soker2026Failed, Soker2026DustJets,  WangShishkinSoker2026} for a partial list of papers from last year). In rare cases, a magnetar might add more energy to the ejecta (e.g., \citealt{Kumar2025}). Still, it occurs after the explosion (e.g., \citealt{Nicholletal2017}), and, in most cases with energetic magnetars, jets must drive the explosion because the neutrino-driven mechanism cannot supply the required explosion energy (e.g., \citealt{Kumar2025}). Although neutrino heating can boost the jets \citep{Soker2022nu}, jets supply most of the explosion energy. 
     
The competing delayed neutrino-driven (neutrino-heating) mechanism attributes most CCSN explosions to neutrino heating with no role to jets (e.g., some papers from 2026: \citealt{Akaho2026, Akhmetalietal2026, Burrowsetal2026, Calvertetal2026, ChenCHetal2026, EggenbergerAndersenetal2026, Giudicietal2026, LuoZhaKajino2026, Mezzacappa2026, Murphyetal2026, Neopaneetal2026, Orlando2026, PanLi2026, Paradisoetal2026, Rusakovetal2026, vanBaaletal26, VarmaMuller2026, Wessonetal2026}). The neutrino-driven mechanism cannot explain the energetics of CCSNe with explosion energies $\gtrsim 2 \times 10^{51} \erg$, so they are attributed to the magnetorotational explosion mechanism (e.g., \citealt{Jankaetal2016}). The latter occurs only in rare cases, where the pre-collapse core is rapidly rotating (e.g., \citealt{Shibataetal2025, Mannoetal2026, PanLi2026, Griffithsetal2026} for recent papers on the magnetorotational explosion mechanism; or a collapsar in cases where a black hole is formed, e.g., \citealt{BoppGottlieb2025, Gottliebetal2025}).\footnote{Too many researchers confuse the magnetorotational mechanism with the JJEM. However, aside from the fact that both rely on jets to power the explosion, they are significantly different: (1) The JJEM accounts for all CCSNe, including low-energy CCSNe. In contrast, the magnetorotational mechanism operates only rarely and produces energetic explosions. (2) The source of the angular momentum in the magnetorotational mechanism is a rapidly rotating pre-collapse core, while in the JJEM, it is the stochastic convecting motion in the core. These differences lead to large differences in the outcomes of the explosions in the two mechanisms (for a thorough discussion, see \citealt{Soker2025G11}).}   

Studies of CCSN remnant (CCSNR) morphologies in the last decade (starting with \citealt{GrichenerSoker2017}) have revealed many properties of the JJEM, as many of these cannot yet be derived from simulations \citep{Soker2025Learning}. Moreover, CCSNR morphologies are the only type of observable that, at present, can robustly distinguish between the JJEM and the neutrino-driven mechanism (e.g., \citealt{Soker2024UnivReview}). Most significantly are CCSNRs with point-symmetric morphologies  (e.g., review by \citealt{Soker2024Rev}),  which the JJEM predicts for many (but not all) CCSNRs, e.g., by three-dimensional hydrodynamical simulations (e.g., \citealt{Braudoetal2025, Braudoetal2026, AkashiSoker2026a, AkashiSoker2026BG11, BraudoSoker2026}), and the neutrino-driven and the magnetorotational mechanisms have no explanation for 
 (e.g., \citealt{SokerShishkin2025Vela}; for more general comparison of the explosion mechanisms see table 2 in \citealt{Soker2025Learning}, and table 1 in \citealt{Soker2025G11}).  Studies in the past five years identified about 20 CCSNRs with point-symmetric morphologies attributed to the JJEM, starting with SNR 0540-69.3 (\citealt{Soker2022SNR0540}; confirmed by \citealt{ShishkinSoker2026SNR0540}), and the last addition of SNR G7.7-3.7, where \cite{Luoetal2026G77} identified two pairs of opposite ears (which they termed blowouts) even in this CCSNR with a highly distorted morphology. I attribute this structure to the JJEM.

One open question in JJEM concerns the number of jet pairs participating in the explosion, the duration of the jet-launching episodes, and the jets' energies. I elaborate on this and on the motivation to conduct this study in Section \ref{sec:Motivation}. As with many other studies in the JJEM, observations motivate this study, and I describe them there. Some CCSNRs have morphological signatures of more than five pairs of jets, indicating several intermittent accretion disk episodes. Some other CCSNRs reveal only two or three very energetic jet pairs, indicating long-lived intermittent accretion disks. In Section \ref{sec:Disk}, I propose an explanation for these long-lived intermittent accretion disks. 
I summarize this study in Section \ref{sec:Summary}. 

\section{Motivation}
\label{sec:Motivation}

 The Crab Nebula CCSNR has a rich point-symmetric morphology. Based on observations, \cite{ShishkinSoker2025Crab} first identified its rich point-symmetric morphology in JWST IR observations: seven pairs of `bays', and one pair of filaments (\citealt{Blairetal2026} rediscovered the pair of filaments). \cite{Temimetal2024} marked nine of the bays (the pair of the largest two bays was identified much earlier, e.g., \citealt{Micheletal1991, Fesenetal1992}), but they did not notice the point-symmetric morphology. Simulations of the JJEM (Section \ref{sec:intro}) show that bays and the filaments are not necessarily along axes of jet pairs. \cite{GrichenerSoker2017} considered the long axis of the Crab Nebula to be a jet axis in the framework of the JJEM based on the southeast ear (a protrusion from the main nebula). However, there is no observable ear on the other side. \cite{Dingetal2026} studied the northern jet of the Crab Nebula, but argue that there is no counterjet. I identified a southern ring opposite to the northern jet, which I attributed to a counterjet \citep{Soker2026RNAAS}. This is a robust jet axis of the crab. The nine identified symmetry axes of the Crab Nebula suggest that five or more pairs of jets exploded this CCSNR. The low explosion energy of the Crab Nebula, $E_{\rm exp} \simeq 5 \times 10^{49} -10^{50} \erg$ (e.g., \citealt{YangChevalier2015}; note that \citealt{BietenholzNugent2015} claimed a somewhat larger energy), implies low-energy pairs of jets.
 (For the properties of the jets in the JJEM, see Table 1 in \citealt{Soker2025Learning}.)   

SNR G7.7-3.7 has two pairs of small ears that \cite{Luoetal2026G77} identified. This indicates two pairs of jets that comprise a small fraction of the total explosion energy. There are no morphological indications for very energetic pairs of jets. The same holds for the Vela SNR, with its rich point-symmetric morphology (e.g., \citealt{SokerShishkin2025Vela}), indicating several pairs of jets, but none that is very energetic.  
 The energy that a jet carries is crudely estimated by the volume of the ear or lobe the jet inflates. \cite{GrichenerSoker2017} applied this method first and estimated the jets' energies by the relative volume of an ear in a CCSNR to the total volume of the CCSNR. They somewhat underestimated the jet energy because they did not account for the material the jets shaped inside the main CCSNR shell; instead, they measured only the volume of the jet-shaped ear protruding from the main CCSNR shell. The same general relative-volume method is used to estimate the energy the jet carried to inflate a structure. 

On the other hand, some CCSNR morphologies exhibit two or three pairs of jets, which account for most of the explosion energy within the JJEM framework.    
Examples include SNR S147 \cite{Shishkinetal2025S147}, SNR W44 \citep{Soker2024W44}, and SNR RCW 89 \citep{Soker2025RCW89}, each having two prominent jet axes. 
SNR G0.9+0.1 has three clear symmetry axes, with two pairs of jets launched along the long axis \citep{Soker2025G0901}. Namely, it was shaped by at least two regular-energy jet pairs and two energetic jet pairs. 
The presence of two or more pairs of jets with misaligned axes implies that, in these cases, the explosions are not due to the magnetorotational mechanism, despite the energetic jet pairs. In this study, I address the formation of energetic jet pairs.  

For SNR Puppis A, \cite{Bearetal2025Puppis} identified three pairs of structural features indicating regular-energy pairs, but also proposed one very energetic pair with one jet much more powerful than the opposite one. They proposed that the NS launched this pair early in the explosion process, when the mass accretion rate onto the newly born NS was high. They proposed that the asymmetrical pair of jets imparted the natal kick to the NS in what they proposed and termed the kick-BEAP: kick by early asymmetrical pair. 
However, not all energetic jet pairs are launched early in the explosion process.\footnote{In identifying the signatures of energetic jets in CCSNRs, the papers cited above mainly employed methods commonly used in the study of planetary nebulae. I describe these in the Appendix.}   

In \cite{Soker2025Learning} I listed the jet-launching episode timescales to be $\tau_{\rm 2j} \simeq 0.01-0.3 \s$;  
 the typical lifetimes of the intermittent accretion disks, $\tau_d$, are about the same. This is estimated from the explosion timescale of one to a few seconds, and the estimated number of exploding pairs of jets, $\simeq 5-30$ calculated from the estimated energy of each typical pair of jets \citep{Soker2025Learning}.   
The upper limit of $\simeq 0.1-0.3 \s$ is based on observations of energetic jets. The expectation from convective fluctuations alone is a disk lifetime of $\tau_{\rm d} \lesssim 0.1 \s$. For example, \cite{GilkisSoker2015} estimated the number of convective cells in a shell accreted from the core at a given time to be $\simeq 20-40$ convective cells. This gives a short fluctuation time. Also, in \cite{Soker2018arXiv}, I analyze the results of simulations by \cite{Mulleretal2017}, where the angular momentum fluctuation lifetimes are $\lesssim 0.1 \s$. Long-lived, intermittent accretion disks require a mechanism to prolong their lifetimes.  

Since the expectation is that there will be $\simeq 5-20$ pairs of jets, the existence of two or three very energetic pairs of jets (lower-energy pairs of jets are possible, and even likely) implies long-lived intermittent accretion disks. I aim to explain the existence of such long-lived intermittent accretion disks. 
   
In \cite{Soker2025RCW89}, I proposed a process that is based on feedback from the jets. I proposed that two energetic jets of a pair might inflate large bubbles as they interact with the stellar core material. The bubbles prevent accretion from their direction; hence, accretion continues from the equatorial plane between them. The angular momentum of this material, which flows towards the NS in the equatorial plane, is along the same direction as that of the pair of jets that inflated the pair of bubbles, or opposite to it. The newly accreted material continues to maintain the original accretion disk, or the opposite one.  In this study, I focus on a different mechanism that is based on disk viscosity.  
I turn to study this process that can prolong the lifetime of intermittent accretion disks. The two processes likely act together to prolong the lifetime of intermittent accretion disks. 

\section{Long-lived intermittent disks}
\label{sec:Disk}
\subsection{Disk properties}
\label{subsec:DiskProperties}

I consider intermittent accretion disks around the newly born NS. Because each disk lasts for a short time of $\tau_d \simeq 0.01-0.1 \s$ and the viscosity relaxation time is $\tau_{\rm vis} \simeq 0.01-0.1 \s$ \citep{Soker2025Learning}, it might not fully relax during its lifetime. The entire vicinity of the disk is opaque to photons, and most of the disk is optically thin to neutrinos. For these three reasons, I cannot use the standard model for geometrically thin, optically thick accretion disks. I will use the $\alpha$ prescription for the disk viscosity.  Although the disk might not be relaxed, I use the same viscosity timescale. This is a common assumption when studying the role of viscosity in accretion disks: the same viscous timescale is used whether the disk is relaxed or not (e.g., \citealt{FrankKingRaine1985Book}). Also, although intermittent accretion disks in the JJEM often have no time to fully relax by viscosity and possibly the thermal timescale, the disk lives much longer than the dynamical (Keplerian) timescale. Therefore, they can be treated as an accretion disk rather than a parcel of material that is accreted from a specific direction.  

I start, therefore, with the viscosity timescale I estimated in \citep{Soker2025Learning}, and the expression (e.g., \citealt{FrankKingRaine1985Book}) 
\begin{equation}
  \tau_{\rm vis} \simeq \frac{1}{\alpha}  \left(\frac{R}{H}\right)^2 \frac{R}{v_{\rm K}} 
  = \frac{1}{2 \pi \alpha}  \left(\frac{R}{H}\right)^2 \tau_{\rm K} , 
    \label{eq:tvis1}
\end{equation}
where $\alpha$ is the usual viscosity parameter, $\nu = \alpha  C_s H$, $C_s$ is the sound speed in the disk, H the scaleheight at radius $R$, and 
\begin{equation}
  v_{\rm K} = 6.10 \times 10^4
  \left( \frac{M_{\rm NS}}{1.4 M_\odot} \right)^{1/2}
  \left( \frac{R}{50 \km} \right)^{-1/2}
  \km \s^{-1}, 
    \label{eq:vkep}
\end{equation}
the Keplerian velocity at radius $R$ and for an NS mass of $M_{\rm NS}$. I scale with the radius corresponding to the outer regions of intermittent accretion disks around the newly born NS.  
The Keplerian period is
\begin{equation}
  \tau_{\rm K} = 0.00515
  \left( \frac{M_{\rm NS}}{1.4 M_\odot} \right)^{-1/2}
  \left( \frac{R}{50 \km} \right)^{3/2}   \s. 
    \label{eq:taukep}
\end{equation} 
Using the typical parameters cited above, I write the expression for the viscosity parameter
\begin{equation}
\alpha \simeq 0.15 
\left( \frac{\tau_{\rm vis}}{0.05 \s} \right)^{-1}
\left( \frac{\tau_{\rm K}}{0.00515\s} \right)
\left( \frac{R}{3H} \right)^2 .
    \label{eq:alpha}
\end{equation}
The inward radial velocity of the material is (e.g., \citealt{KotkoLasota2012}) 
\begin{equation}
v_{R} \simeq \frac{R}{t_{\rm vis}} = 1000 
\left( \frac{R}{50 \km} \right)
\left( \frac{\tau_{\rm vis}}{0.05 \s} \right)^{-1}
\km \s^{-1}. 
    \label{eq:vR}
\end{equation}

The mass accretion rate is 
\begin{equation}
\dot M_{\rm acc}=2H 2 \pi R \rho v_{R} = 2 \pi R \Sigma v_R, 
    \label{eq:Macc}
\end{equation}
where $\rho$ is the appropriate average density in the disk, and $\Sigma=2 H \rho$ is the surface density. The density is 
\begin{equation}
\begin{split}
\rho \simeq & 5.7 \times 10^{10}  
\left( \frac{\dot M_{\rm acc}}{0.3 M_\odot \s^{-1}} \right)
\left( \frac{R}{50 \km} \right)^{-2}
\\ & \times 
\left( \frac{R}{3H} \right)
\left( \frac{v_{R}}{1000 \km \s^{-1}} \right)^{-1}
\g \cm^{-3}.
    \label{eq:rho}
\end{split}
\end{equation}
I scaled by $\dot M_{\rm acc} = 0.3 M_\odot \s^{-1}$ according to \cite{Mulleretal2017}, but note that for $t \lesssim 0.25 \s$ the mass accretion rate can be higher, and it decreases to $\simeq 0.1 M_\odot \s^{-1}$ at $\simeq 0.6 \s$. 

I derived these parameter values to present the reader with the general order-of-magnitude properties of intermittent accretion disks in the JJEM and to emphasize that they are based mainly on observations of CCSNRs (and some one-dimensional simulations of pre-collapse stellar models), not on collapse simulations. 

A key issue, as mentioned above and as I reiterate, is that the accretion episode's lifetime, during which the accreted gas maintains its specific angular momentum, is short. Namely, it lasts for about the viscosity timescale, or shorter, $\lesssim \tau_{\rm vis}$.  I examine a way to prolong the lifetime of the accretion disk beyond the lifetime of the accretion episode that formed the temporary accretion disk. And again, each intermittent accretion disk is unlikely to be fully relaxed due to its short lifetime. 

\subsection{Prolonging disk lifetime by outward angular momentum transport}
\label{subsec:Prolonging}

I consider a simple, idealized accretion flow evolution within the framework of the JJEM, i.e., where fluctuations in the specific angular momentum of the accreted gas persist, even though the average angular momentum is zero over time. 
Although the accretion is continuous, I construct it from idealized accretion episodes, one after the other. 

An accretion episode with large-angular-momentum material, namely, a large fluctuation in the angular momentum, forms an accretion disk. This accretion episode lasts for $\gtrsim \tau_{\rm vis} \simeq 0.05-0.1 \s$ to build a developed disk. 
This accretion disk launches two opposite jets. The accretion disk lifetime from the material of this accretion episode alone should last for $\lesssim 0.1 \s$. Immediately after that, the newly accreted material has zero angular momentum. 

\begin{figure} 
\centering
\includegraphics[trim=0.8cm 19.0cm 0.0cm 0.3cm ,clip, angle=0, scale=0.50]{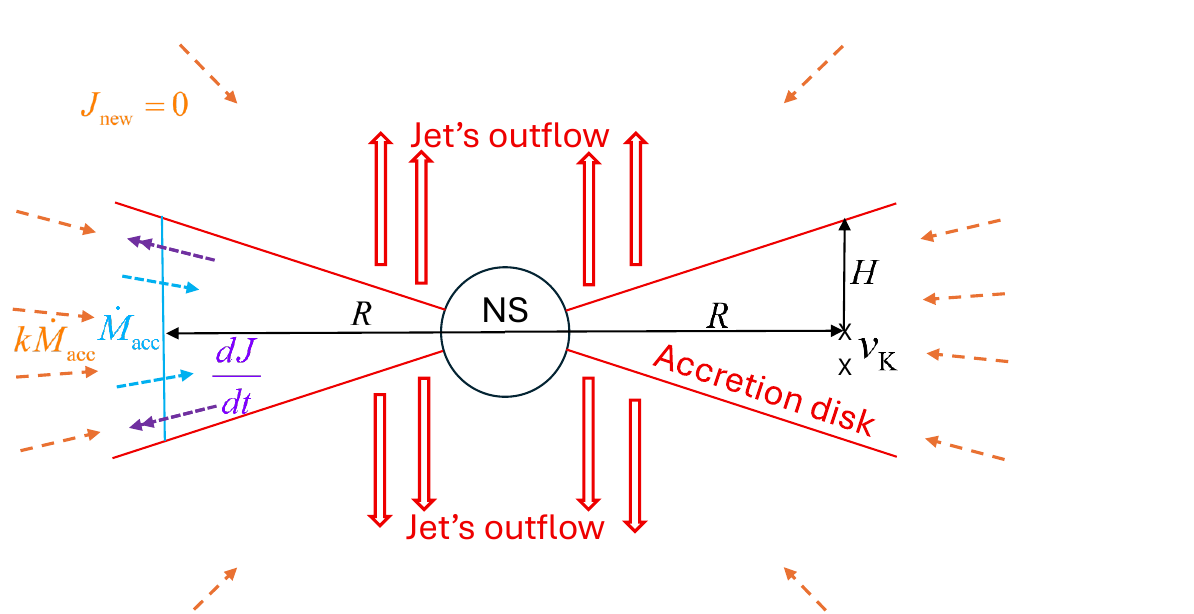}
\caption{A schematic drawing of the flow structure when a new accretion episode with an average zero angular momentum occurs immediately after an accretion episode that formed a developed and more or less relaxed accretion disk that launches jets (depicted by the red double-lined arrows). 
The NS is at the center, and the disk's angular momentum is vertical in the figure. The equatorial plane is horizontal along the two arrows that mark the radius $R$ towards an arbitrary radius. The two 'x's stand for a Keplerian velocity into the page on the right side, and out of the page on the left side. The mass accretion rate through the disk in the early mass accretion episode that formed the accretion disk is $\dot M_{\rm acc}$ (depicted by the dashed-pale-blue arrows). The new accretion episode inflows mainly from the vicinity of the equatorial plane because the two jets prevent inflow from the polar directions; only a fraction $k$ of the accretion inflow at large radii (before the encounter with the jets) reaches the accretion disk.  
The viscosity transfers angular momentum outwards at a rate of $dJ/dt$ (depicted by dashed-double-headed purple arrows). 
}
\label{fig:Schematic}
\end{figure}

The newly accreted material adds mass with zero angular momentum to the accretion disk, which reduces the average specific angular momentum of the accretion disk. Material loses specific angular momentum, hence centrifugal support, and the disk in its inner parts shrinks and accretes onto the NS. However, further out, where I assume the disk does not launch the jets that carry angular momentum, the situation is different. 
I schematically draw the flow structure in the outer disk in Figure \ref{fig:Schematic}. 

The viscosity in the accretion disk transfers angular momentum outwards at a rate of (e.g., \citealt{FrankKingRaine1985Book})
\begin{equation}
\dot J_{\rm out} \equiv \left(\frac{dJ}{dt}\right)_R = -2 \pi R \nu \Sigma R^2 \frac{d \Omega} {dR} = 3 \pi \nu \Sigma R v_{\rm K},  
    \label{eq:dJdt1}
\end{equation}
where $\Sigma=2H \rho$ is the surface density at radius $R$, and in the second equality I assumed a Keplerian accretion disk ($\Omega = \sqrt{GM/R^3}$). I approximate the accretion rate onto the disk by the solid angle of the outer edge of the disk $\simeq 4 \pi H/R$; namely, only a fraction of $k \simeq (H/R)$ reaches the outer accretion disk. After some time, the accretion rate in the disk would be 
\begin{equation}
\dot M_{\rm out,new}=k \dot M_{\rm acc} \simeq \frac {H}{R} \dot M_{\rm acc}.
    \label{eq:Moutnew}
\end{equation}
In any case, I take the accretion rate onto the outer accretion disk to be $k$ times the disk accretion rate. When the accretion rate and its angular momentum increase with time, it is possible that $k>1$. I do not consider these cases in the simple flow structure I study here. 
In a steady-state accretion disk, the rate of angular momentum that the viscosity transfers outwards divided by the mass accretion rate equals the specific angular momentum at that radius. In a case where the accretion rate changes by a factor $k$ as defined above, the expression is 
\begin{equation}
j_{\rm \nu,out} \equiv \frac{\dot J_{\rm out} }{\dot M_{\rm out,new}} \simeq \frac {1}{k} j_{\rm K}. 
\label{eq:jnuout}
\end{equation}

In the transition from the first to the second mass accretion episodes I study here, $k<1$, and the specific angular momentum of the material in the outer disk increases, even though the newly accreted material has zero angular momentum. Later, the accretion rate settles to $\dot M_{\rm out,new}$, and  $j_{\rm \nu,out}=j_{\rm K}$. In principle, the disk can live forever as angular momentum flows out, and material with zero angular momentum is accreted through the disk and onto the NS. This is, of course, not accurate, as the central object, here an NS, rotates and its radius is non-negligible, so the final angular momentum of the accreted gas is not zero and is non-negligible.  The jets also carry away angular momentum. 

I take a different approach. I assume that the jets carry angular momentum from some inner radius in the disk, close to the NS, $R_{\rm d,j} \simeq 20 \km$. When the transition from the first to the second accretion episodes occurs, the outer radius of the disk is $R_{\rm D} \simeq 50-60 \km$. I also assume that the surface density in the disk is that of a blackbody-emitting disk (although this is not the case here), $\Sigma \propto r^{-3/4}$ (e.g., \citealt{FrankKingRaine1985Book}).
The total angular momentum of the disk at the transition between the two accretion episodes is 
\begin{equation}
\begin{split}
J_{\rm D,0} & =\int^{R_{\rm D}}_{R_{\rm d,j}} 2 \pi r dr \Sigma \sqrt{G M_{\rm NS} r} = 
M_{\rm D,0} j_{\rm d,j} 
\\ &\times 
 \left[ \frac{5}{7}\frac{(R_{\rm D}/R_{\rm d,j})^{7/4} - 1}{(R_{\rm D}/R_{\rm d,j})^{5/4} - 1}  \right] \equiv M_{\rm D,0} j_{\rm d,j}  Q,
    \label{eq:JDisk}
\end{split}
\end{equation}
where $M_{\rm D,0}$ is the accretion disk mass at that time, and $j_{\rm d,j}=\sqrt{G M_{\rm NS} R_{\rm d,j}}$ is the specific angular momentum in the inner disk region that launches the jets. The third equality defines $Q$ to be the factor inside the square parenthesis: $Q=1.41$ for $R_{\rm D}=3 R_{\rm d,j}$, $Q=1.32$ for $R_{\rm D}=2.5 R_{\rm d,j}$, and $Q=1$ for $R_{\rm D} \rightarrow R_{\rm d,j}$.
The accretion disk will survive as long as the specific angular momentum of the disk is $>j_{\rm d,j}$. 

Let an extra mass, 
\begin{equation}
\Delta M_{\rm D}=k\dot M_{\rm acc} \Delta t_{\rm D},
 \label{eq:DeltaMDef}
\end{equation}
(by equation  \ref{eq:Moutnew}) be accreted through the disk in the second episode, which has zero angular momentum, over a time of $\Delta t_{\rm D}$ until the disk ceases to exist. 
The total angular momentum that is lost in the inner radius, to the accreted mass onto the NS and to the jets, is $(M_{\rm D,0} + \Delta M_{\rm D}) j_{\rm d,j}$. This equals the angular momentum of the disk at the transition between the two episodes, $J_{\rm D,0}$, because I assume here that the material in the second episode brings more mass.  
From this equality, $(M_{\rm D,0} + \Delta M_{\rm D}) j_{\rm d,j}=J_{\rm D,0}$, and using equation (\ref{eq:JDisk}), I find that the extra mass that is accreted through the disk is 
\begin{equation}
\Delta M_{\rm D} \simeq 0.3 M_{\rm D,0} \frac{Q-1}{0.3}.
    \label{eq:DeltaM}
\end{equation}
The duration of the second accretion episode while the disk exists is given by equation (\ref{eq:DeltaMDef}). Substituting $k=H/R$ as in equation (\ref{eq:Moutnew}), $\dot M_{\rm acc}=M_{\rm D,0}/\tau_{\rm vis}$ during the first accretion episode, and using equation (\ref{eq:DeltaM}), gives.   
\begin{equation}
\Delta t_{\rm D} \simeq \tau_{\rm vis}  \frac{Q-1}{0.3} \frac{R}{3H}. 
    \label{eq:Deltat}  
\end{equation}
 I take the fraction of the accreted gas that is accreted through the disk as $k \simeq (H/R)$, because at this phase I assume a low specific angular momentum of the accreted gas. On average, accretion occurs via radial inflow. The solid angle that the outer surface of the disk covers is $4 \pi \sin \alpha_d = 4 \pi H/(H^2+R^2)^{1/2} \simeq 4 \pi H/R $, where $\alpha_d$ is the half-opening angle of the disk, and I assume H<<R. The rest of the mass falls directly to the central object (the newly born NS).  

 I crudely estimate a plausible value for $\Delta t_{\rm D}$. I consider a disk that has relaxed somewhat and is therefore relatively thin, $H\simeq 0.2R$. From equations (\ref{eq:tvis1}) and (\ref{eq:taukep}) with $R\simeq 50 \km$, I find $\tau_{\rm vis} \simeq 0.2 (0.1/\alpha) \s$. With $Q \simeq 1.3$ (as mentioned above), I find from equation (\ref{eq:Deltat}) that $\Delta t_{\rm D} \simeq 0.3 \s$. This is the extra lifetime of the accretion disk; the total lifetime is somewhat longer. Therefore, the value of $\simeq 0.3 \s$ that I mentioned earlier for the disk lifetime is plausible under favorable conditions. In the JJEM, most pairs of jets are active for much shorter periods. But in some CCSNe, 1-3 pairs of jets might be active for such a long time. 

The conclusion from this simple derivation is that the disk can survive for a typical time of the order of its viscous time when accreting material with zero angular momentum (equation \ref{eq:Deltat}). The same holds for an accretion episode with angular momentum fluctuations around zero with typical variation time much shorter than $\tau_{\rm vis}$. The amount of mass accreted is a fraction of the disk mass, as given by equation (\ref{eq:DeltaM}). In other words, the end of an accretion episode with large specific angular momentum does not necessarily imply that the accretion disk ceases to exist. 

\subsection{Combining the different effects}
\label{subsec:Combining}

I extend the discussion to include several accretion episodes after the first one studied here (which need not be the first one in the entire explosion process),  and processes studied in earlier papers.   
   
There is an effect where the jets prevent accretion from their directions of propagation \citep{PapishSoker2014Plan, Soker2025RCW89}, i.e., the polar directions and vicinity. If the jets inflate bubbles, the accretion will be limited to the vicinity of the equatorial plane. This gas has its angular momentum aligned with the original jets as it flows in the jets' equatorial plane. The fluctuations of angular momentum will be either along or opposite to the angular momentum of the jets of the first episode. This `random walk' of angular momentum fluctuations might result in a total angular momentum along the original one. This further prolongs the accretion disk's lifetime beyond the one I calculated here, which was for a sum of fluctuations equal to zero after the first episode. The outward transport of angular momentum that I studied above ensures the continuation of the jet activity.
The total angular momentum of the core is zero (or very small). The sum of the fluctuations that results in a positive value of the gas accreted onto the NS implies that gas with a negative angular momentum was expelled by the jets and the bubbles they inflated.    

In some cases, the accretion episode that follows the one that formed the disk proceeds in the opposite direction, and the disk lives for only a very short time after the transition between the two accretion episodes. In others, as I described above, the disk's lifetime can be long, and it launches two opposite jets that together carry a large fraction of the explosion energy. The key result is that processes can prolong the accretion disk's lifetime once it develops: the jets it launches, particularly if they inflate bubbles, and the disk's viscosity, which transports angular momentum outwards as I showed in Section \ref{subsec:Prolonging}.

\section{Summary} 
\label{sec:Summary}

Motivated directly by observations of CCSNRs that suggest cases where two or three energetic pairs of jets dominate the CCSNR morphology, and hence its explosion power, I examine the formation of long-lived intermittent accretion disks in the framework of the JJEM. These long-lived intermittent accretion disks launch the energetic jets. In addition to these energetic pairs, the JJEM predicts cases where several more pairs of weak jets also participate in the explosion. According to the JJEM, in some cases only weak jets can explode the star, with up to 20 (or even 30) jet pairs.   

Three processes can act together to form such long-lived intermittent accretion disks: two are positive feedback processes, and one is stochastic (summarized in Section \ref{subsec:Combining}).  In this paper, I calculated only the role of viscosity (second below). The other two effects come from earlier papers.  

(1) \textit{Forcing accretion from the equatorial plane vicinity} \citep{Soker2025RCW89}. This positive feedback effect operates when the disk launches powerful jets. The jets launched by the accretion disk interact with the infalling core material at distances of hundreds to thousands of km and prevent accretion along and near the polar direction. The accretion proceeds from the vicinity of the equatorial plane. The gas that flows inward in the vicinity of the equatorial plane has its angular momentum either (more or less) along the angular momentum of the disk or opposite to it. On average, it can increase the fluctuations of the specific angular momentum along the disk's angular momentum.  It operates in a positive feedback cycle because the jets might prolong their life. 

(2) \textit{Prolonging the disk lifetime by viscous angular momentum transfer.}
This process was the subject of this study. 
In this study, I examine a typical intermittent accretion disk (Section \ref{subsec:DiskProperties}) that begins accreting mass with zero angular momentum (Section \ref{subsec:Prolonging}). The disk forms during an accretion episode with positive angular momentum and becomes well developed; i.e., the gas settles near the equatorial plane so that the angular velocity decreases outward, becoming Keplerian. In the simple flow I considered, the first accretion episode is followed by a second one in which the accreted gas has zero angular momentum, or by angular-momentum fluctuations on short timescales that cancel each other. I found that in this simple flow evolution, viscosity might allow the disk to live to about its viscous timescale (equation \ref{eq:Deltat}). Namely, the disk does not cease to exist when the material with zero angular momentum starts to be accreted. This adds more mass to the disk (equation \ref{eq:DeltaM}). Namely, even though the accreted mass in the second episode has zero angular momentum, it is still accreted from an accretion disk, at least at the beginning of that accretion episode.    
 This process operates in a positive feedback cycle because, for the mechanism to work, the disk should be relatively relaxed with $H \ll R$ (equation \ref{eq:Deltat}). Namely, the disk already lives for a long time, and this allows it to prolong its life.

(3) \textit{Random fluctuations.} The specific angular momentum of material that is accreted onto the already developed accretion disk is likely to fluctuate around zero. In about half of the cases, this `random walk' of fluctuations sums to a positive value (with the same sign as the original accretion disk). Over a not-too-long time, the sum might be non-negligible relative to the minimum needed to maintain the disk, $j_{\rm d,j}$, hence extending the lifetime of the accretion disk. This will not happen too often. It might operate once, twice, three times, or not at all. 

I suggest that these three effects can add together, particularly the first two, which are positive feedback cycles, and lead to long-lived, very energetic jet pairs.  I tentatively estimate that the first two processes are of the same order of magnitude, and dominate in most cases. Only in rare cases might the third, stochastic process dominate. 

For now, my suggestion is somewhat speculative. Future hydrodynamical simulations will have to examine all three processes. Such simulations will be extremely resource-intensive. However, observations of CCSNRs tend to support the JJEM, with some CCSNRs indicating two or three energetic pairs of jets. To better understand the explosion mechanism of massive stars in CCSNe, the community will need to invest substantial resources in such simulations.     


\bibliography{reference-file.bib}


\section{ACKNOWLEDGEMENTS}

 I thank an anonymous referee for detailed and thorough comments and suggestions that substantially improved the manuscript. 
I thank the Charles Wolfson Academic Chair at the Technion for the support.

\section*{APPENDIX}

I briefly describe the method used to establish the major role of jets in planetary nebulae, which has been important for the study of CCSNRs over the last decade, particularly point-symmetric CCSNRs in the last five years.  

The claim for two or three pairs of energetic jets in CCSNRs is based on the identification of large jet-shaped morphologies in several CCSNRs. Although this method of point-symmetric morphology identification is new (only about five years old) in he study of CCSNRs, it is a mature and well developed method in the study of planetary nebulae  (e.g.,  \citealt{Morris1987, Soker1990AJ, SahaiTrauger1998, AkashiSoker2018,   EstrellaTrujilloetal2019, Tafoyaetal2019, Balicketal2020,   GarciaSeguraetal2020, GarciaSeguraetal2021, Clairmontetal2022, RechyGarciaetal2020, Danehkar2022, MoragaBaezetal2023, Ablimit2024, Derlopaetal2024, Mirandaetal2024, Sahaietal2024, Kwoketal2026, Masaetal2026}, for a very partial list). These and many other studies have argued that point-symmetric planetary nebulae testify to the major roles of jets in shaping and powering planetary nebula outflows. 
Signatures include opposite (to the center) structural features of lobes, cavities, ears, and rings. Many planetary nebulae show signatures of powering and shaping by precessing jets (e.g., \citealt{Guerreroetal1998, Mirandaetal1998, Sahaietal2005, Boffinetal2012, Sowickaetal2017, RechyGarciaetal2019, Guerreoetal2021, Clairmontetal2022}). 

The main method that established the major role of jets in planetary nebulae began with careful visual inspection and qualitative morphological classifications (e.g., \citealt{Balick1987, Parkeretal2006, Sahaietal2007, Kwok2024}). Although qualitative, this method enabled robust identification of a wide variety of jet-shaped structures (e.g., \citealt{SahaiTrauger1998}). This, in turn, motivated relevant numerical simulations, which shed light on the shaping processes through qualitative comparisons with observations (e.g., \citealt{GarciaSeguraetal2021, GarciaSeguraetal2022, GarciaSeguraetal2025}). For a new quantitative method for identifying point-symmetric morphologies, see \citealt{ShishkinMichaelis2026}.)

\end{document}